\newcommand\aap{A\&A}
\documentclass{iaus}
\usepackage{graphicx}

\title[Loud AGN with {\it XMM-Newton}] %% give here short title %%
{Radio Loud AGN in the 2XMMi catalogue}

\author[Labiano, Guainazzi, Bianchi, Miniutti]   %% give here short author list %%
{Alvaro Labiano$^1$,
      Matteo Guainazzi$^1$,
   Stefano Bianchi$^2$}

\affiliation{$^1$European Space Astronomy Centre of ESA, Madrid, Spain \\ email: {\tt Alvaro.Labiano@esa.int, Matteo.Guainazzi@esa.int} \\[\affilskip]
$^2$Dipartimento di Fisica, Universit\`a degli Studi Roma Tre, Rome, Italy \\ email: {bianchi@fis.uniroma3.it}\\[\affilskip]
}

\pubyear{2010}
\volume{xxx}  %% insert here IAU Symposium No.
\pagerange{0000}
\jname{IAU Symposium 275: Jets at all Scales }
\editors{Gustavo E. Romero, Rashid A. Sunyaev \& Tomaso Belloni, eds.}
\begin{document}

\maketitle

\begin{abstract}
We are carrying out a search for all radio loud Active Galactic Nuclei observed with {\it XMM-Newton}, including targeted and field sources to perform a multi-wavelength study of these objects. We have cross-correlated the Ver\'on-Cetty \& Ver\'on (2010) catalogue with the {\it XMM-Newton} Serendipitous Source Catalogue (2XMMi) and found around 4000 sources. A literature search provided radio, optical, and X-ray data for 403 sources. This poster summarizes the first results of our study.
\end{abstract}

\firstsection % if your document starts with a section,
\section{Introduction and Sample}

\begin{figure}[b]
% \vspace*{-2.0 cm}
\begin{center}
 \includegraphics[width=3.4in]{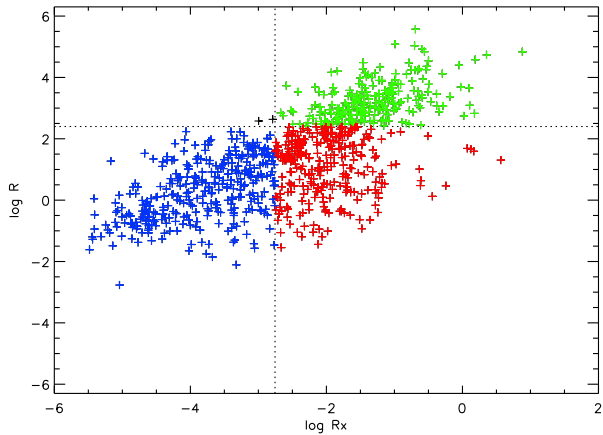} 
% \vspace*{-1.0 cm}
 \caption{Radio loud (green) and radio quiet (blue and red) AGN. Rx is the
ratio between 5 GHz and 2-10 keV emission. R is the ratio between 5
GHz and B-band emission. Boundares from Panessa et al. 2007. Blue points are sources classified as radio quiet by R and Rx parameters  Red points are sources classified as radio quiet by the R parameter and radio loud by the Rx parameter }
   \label{loudquiet}
\end{center}
\end{figure}

\begin{figure}[h]
% \vspace*{-2.0 cm}
\begin{center}
 \includegraphics[width=3.4in]{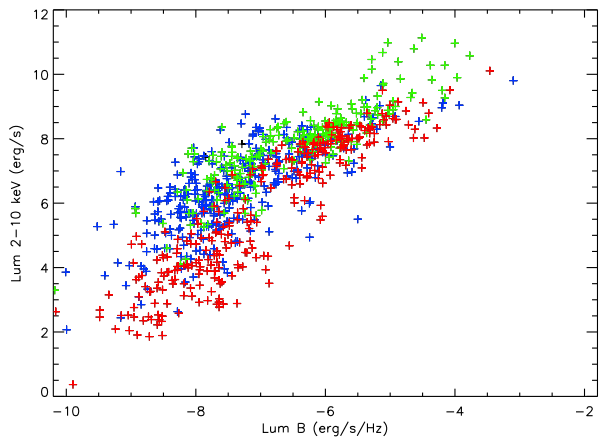} 
% \vspace*{-1.0 cm}
 \caption{X-ray versus B-band luminosities. For
optical luminosities higher than 10$^{-6}$ erg/s/Hz radio loud (green) AGN are brighter in X-rays than radio quiet (red and blue). This effect is stronger at higher luminosities (10$^{-5}$-10$^{-4}$ erg/s/Hz), where radio loud AGN deviate from the low luminosities correlation. X-ray emission in radio loud sources could have larger contributions from the jet.}
   \label{lxlb}
\end{center}
\end{figure}

Bianchi et al. (2009a,b) presented the Catalogue of Active Galactic Nuclei (AGN) in the {\it XMM-Newton} Archive (CAIXA). They focused on the radio-quiet, X-ray unobscured (NH $< 2\times\ 10^{22}$ cm$^{-2}$) AGN observed by XMM-Newton in targeted observations. We are carrying out a similar multiwavelength study, for both targeted and field radio-loud AGN  observed by {\it XMM Newton}. We cross-correlated the Ver\'on-Cetty \& Ver\'on (2010) catalogue (Quasars and Active Galactic Nuclei, 13th edition) with the  {\it XMM-Newton} Serendipitous Source Catalogue (2XMMi, Watson et al., 2009) Third Data Release, and obtained a list of around 4000 sources. However, only  10\% of the sources have published optical and radio data. Our sample consists of all AGN (403 total, Figures \ref{loudquiet} and \ref{lxlb}) with available X-ray (2-10 keV), optical (B-band) and radio (5 GHz) data.

\section{First results and ongoing work}

%Our sample consists of 403 sources with radio, optical and X-ray data, 20\% loud, 80\% quiet AGN (Figure \ref{lxlb}). 
Radio loud sources show jet contribution to optical and X-ray emission, and are brighter in X-rays than radio quiet. Optical and X-rays are AGN dominated with small contribution from host. For optical luminosities higher than 10$^{-6}$ erg/s/Hz radio loud AGN are brighter in X-rays than radio quiet. This effect increases for higher luminosities ($10^{-5}-10^{-4}$), where loud AGN deviate from the low luminosities correlation. X-rays in radio loud sources could have higher contributions from the jet.
%There is a fast increase in X-ray emission in the optically-brightest loud AGN. 
The sample seems to be missing faint, radio loud AGN, although at this point it is not clear if this is due to selection  or astrophysical effects.

While X-rays in radio loud AGN seem to come mainly from jets, other mechanisms of X-ray emission are being studied (e.g. ADAF)? A complete spectral optical and X-ray analysis, including also the 0.2-2 kev band will bring light to the origin of the X-rays in these sources. We are currently studying the sample properties according to different classifications (Seyfert and QSO or FR I and FR II morphologies), and will include IR data when available, with the goal of carrying out a systematic analysis in as many wavelengths as possible.

%\bibliographystyle{/home/alabiano/Astronomy/LaTeX/aa}
%\bibliography{/home/alabiano/Astronomy/LaTeX/ALORefs}

\begin{thebibliography}{4}
\expandafter\ifx\csname natexlab\endcsname\relax\def\natexlab#1{#1}\fi

\bibitem[{{Bianchi} {et~al.}(2009{\natexlab{a}}){Bianchi}, {Bonilla},
  {Guainazzi}, {Matt}, \& {Ponti}}]{Bianchi09b}
{Bianchi}, S., {Bonilla}, N.~F., {Guainazzi}, M., {Matt}, G., \& {Ponti}, G.
  2009{\natexlab{a}}, \aap, 501, 915

\bibitem[{{Bianchi} {et~al.}(2009{\natexlab{b}}){Bianchi}, {Guainazzi}, {Matt},
  {Fonseca Bonilla}, \& {Ponti}}]{Bianchi09a}
{Bianchi}, S., {Guainazzi}, M., {Matt}, G., {Fonseca Bonilla}, N., \& {Ponti},
  G. 2009{\natexlab{b}}, \aap, 495, 421

\bibitem[{{Panessa} {et~al.}(2007) {Panessa}, {Barcons}, \& {Bassani}}]{Panessa07}
{Panessa}, F., {Barcons}, X., \& {Bassani} L. et al., 2009, \aa, 467, 519

\bibitem[{{V{\'e}ron-Cetty} \& {V{\'e}ron}(2010)}]{Veron10}
{V{\'e}ron-Cetty}, M. \& {V{\'e}ron}, P. 2010, \aap, 518, A10

\bibitem[{{Watson} {et~al.}(2009){Watson}, {Schr{\"o}der}, {Fyfe}, {Page},
  {Lamer}, {Mateos}, {Pye}, {Sakano}, {Rosen}, {Ballet}, {Barcons}, {Barret},
  {Boller}, {Brunner}, {Brusa}, {Caccianiga}, {Carrera}, {Ceballos}, {Della
  Ceca}, {Denby}, {Denkinson}, {Dupuy}, {Farrell}, {Fraschetti}, {Freyberg},
  {Guillout}, {Hambaryan}, {Maccacaro}, {Mathiesen}, {McMahon}, {Michel},
  {Motch}, {Osborne}, {Page}, {Pakull}, {Pietsch}, {Saxton}, {Schwope},
  {Severgnini}, {Simpson}, {Sironi}, {Stewart}, {Stewart}, {Stobbart}, {Tedds},
  {Warwick}, {Webb}, {West}, {Worrall}, \& {Yuan}}]{Watson09}
{Watson}, M.~G., {Schr{\"o}der}, A.~C., {Fyfe}, D., {et~al.} 2009, \aap, 493,
  339

\end{thebibliography}
%\begin{thebibliography}{}
%\end{thebibliography}{}

\end{document}